# Photoemission Circular Dichroism and Spin Polarization of the Topological Surface States in Ultrathin Bi$_2$Te$_3$ Films

C.-Z. Xu[1], Y. Liu[1,2], R. Yukawa[3], L.-X. Zhang[1], I. Matsuda[3], T. Miller[1], and T.-C. Chiang[1]

[1]Department of Physics and Frederick Seitz Materials Research Laboratory, University of Illinois at Urbana-Champaign, Urbana, Illinois 61801, USA

[2]X-ray science division, Argonne National Lab, 9700 South Cass Avenue, Lemont, Illinois 60439, USA

[3]Institute for Solid State Physics, The University of Tokyo, Kashiwa, Chiba 277-8581, Japan

**Abstract**

Circular dichroism (CD) observed by photoemission, being sensitive to the orbital and spin angular momenta of the electronic states, is a powerful probe of the nontrivial surface states of topological insulators, but the experimental results thus far have eluded a comprehensive description. We report a study of Bi$_2$Te$_3$ films with thicknesses ranging from one quintuple layer (two-dimensional limit) to twelve layers (bulk limit) over a wide range of incident photon energy. The data show complex variations in magnitude and sign reversals, which are nevertheless well described by a theoretical calculation including all three photoemission mechanisms: dipole transition, surface photoemission, and spin-orbit coupling. The results establish the nontrivial connection between the spin-orbit texture and CD.

Topological insulators such as $Bi_2Te_3$ and $Bi_2Se_3$ are characterized by spin-polarized metallic surface states protected by time-reversal symmetry [1–5], which are promising for applications in spintronic devices and quantum computing [6]. Implementation of these materials in device configurations requires the use of thin films for large scale integration. A further benefit of the thin film configuration is greatly suppressed electrical conduction in the bulk [7], which arises naturally as a result of defects and impurities in real materials but can short out the spin-polarized surface conduction channel. When films become sufficiently thin, quantum size effects can influence the spin texture of the system and thus allow property tuning [8,9]. Experimental characterization of the topological states is essential for understanding the interplay of topological order and quantum confinement. A particularly powerful method is angle resolved photoemission spectroscopy (ARPES) using circularly polarized light, which carries an angular momentum that is well suited for probing the coupled spin-orbital angular momenta of the topological electronic states [10]. The difference between the photoemission intensities arising from excitation by oppositely circularly polarized light yields a circular dichroism (CD) signal, and this method has been widely employed for characterizing the magnetic moments of magnetic materials [11]. More recently, this method has been used to study topological insulators and large CD has been observed from their helical Dirac states. Despite a recent surge of interest in this topic [12–17], the underlying physics for the observed CD remains a challenging issue. Our work on the CD of ultrathin films of $Bi_2Te_3$, reported herein, involves a systematic variation of film thickness and photon energy over a wide range, thus establishing a stringent proofing ground for the theory of CD. The resulting understanding facilitates the design of experiments to extract the spin texture from measurements and is essential for the characterization and engineering of thin films of topological insulators.

Previously, Wang *et al.* performed CD measurements on bulk $Bi_2Se_3$ using 6-eV photons and demonstrated that the results were sensitive to the spin polarization of the initial states [12]. By contrast, another study of $Bi_2Se_3$ led to an explanation in terms of the orbital angular momentum of the initial states [14]. Scholz *et al.* working on bulk $Bi_2Te_3$ discovered that the CD depended on the photon energy, thus indicating a final state effect [16]. Another study of $Bi_2Se_3$ thin films by Vidal *et al.* also suggested the importance of the final states [17]. Our comprehensive measurements show a behavior that is quite complex but can be simplified under appropriate experimental conditions based on calculations involving the interference of three photoemission channels.

Films of $Bi_2Te_3$ were grown by molecular beam epitaxy with a Te/Bi flux ratio of about 3 [18,19] onto a freshly prepared Si (111)-(7×7) substrate maintained at 300°C. The film was annealed at 350°C after deposition. The ARPES measurements were performed at the Synchrotron Radiation Center, University of Wisconsin-Madison. All data were taken with the sample at 50 K using a Scienta analyzer at the U9 PGM-VLS beamline. The energy and momentum resolutions were 20 meV and 0.01 Å$^{-1}$, respectively. First-principles calculations were performed using the ABINIT code [20] and the pseudopotential functions constructed by Hartwigsen, Goedecker, and Hutter [21], which has been shown to yield band structure of $Bi_2Te_3$ films in excellent agreement with experiment [19]. The lattice constants are adopted from a previous publication [22].

The experimental photoemission geometry is shown in Fig. 1. Photoelectrons are measured with different emission angles in the *yz*-plane. The sample is oriented as shown in Fig. 1(a) or 1(b) (geometry A or B, respectively); these two orientations are related by a sample rotation of 30°. The $\overline{\Gamma M}$ direction coincides with a mirror plane. As a result, the CD signal is an odd function of $k_y$ for geometry B, but not for geometry A [23]. The data referred to below are taken with geometry A. Figures. 1(c) and 1(d) show ARPES maps of a film of 2 quintuple layers (QL) of $Bi_2Te_3$ film taken with left- and right-circularly polarized (LCP and RCP) light, respectively. The magnitude of CD is defined as

$$CD = \frac{I_{LCP} - I_{RCP}}{I_{LCP} + I_{RCP}}, \qquad (1)$$

where $I_{LCP}$ ($I_{RCP}$) represents the photoemission intensity measured under LCP (RCP). The photoemission intensity distribution is very different for the two spectra, indicating a large CD. For comparison, Fig. 1(e) shows an ARPES map taken with linearly polarized light for the same 2-QL film, where the conduction and valence band edges (CB and VB, respectively) are indicated. The topological surface state (SS) bands form a Dirac cone at the zone center in the bulk limit, but instead there is a small tunneling gap for the 2-QL film. This gap arises from coupling of the states associated with the top surface and the buried interface of the film [8,28]. Figure 1(f) shows an ARPES map for a 12-QL film acquired with the same linear polarization configuration, and the gap vanishes [19,29]. The tunneling gap is only evident for films of thicknesses of 4 QL and below.

A selected set of CD maps is presented in Fig. 2(a) for film thicknesses of 1-4, 6 and 12 QL taken with photon energies of 19, 29, and 55 eV. The results are further summarized in Fig. 2(b) which shows the CD values of the upper branch of the Dirac cone at $k_y = -0.08$ Å$^{-1}$ for the different film thicknesses at various photon energies. For some photon energies (38 and 50 eV) the CD values for the different film thicknesses are fairly close, but large variations and sign reversals occur at 19 and 55 eV. For the 6- and 12-QL films, which are bulk-like, the largest CD values occur near 29 and 55 eV. The sign reversal at 60 eV agrees qualitatively with the experiment by Scholz *et al.* on bulk Bi$_2$Te$_3$ samples [16]. However, the very strong dependence of the CD on film thickness is surprising, which suggests that prior studies of the CD require further analysis and scrutiny.

The normalized spin polarization of the initial state (upper branch at $k_y = -0.08$ Å$^{-1}$) defined by

$$P(k_y) = \frac{1}{\hbar} \sum_i \langle \psi_i(k_y) | s_x \, \text{sgn}(z) | \psi_i(k_y) \rangle , \qquad (2)$$

is computed by using wave functions $\psi$ obtained from first-principles calculations for freestanding films, where sgn($z$) is the sign function, $z = 0$ is at the midpoint of the film, and the summation over $i$ is for the degenerate Kramers pair associated with the two faces of the film [30]. The computation is limited to 1-6 QL for simplicity. The results for different film thicknesses (Fig. 2(d)) reveal that the spin polarization increases as a function of film thickness of 2 QL and up, but the spin polarization for 1 QL has an opposite sign. This behavior is consistent with the measured spin polarization of the Dirac cone in ultrathin films and bulk Bi$_2$Se$_3$ [9,31]. It is interesting to note that the 1-6 QL films have very similar CD at 50 eV (Fig. 2(b)) even though the spin polarization is very different.

To understand the complex CD behavior, we have performed a calculation with results shown in Fig. 2(c) for comparison with the experiment. The interaction Hamiltonian of an electromagnetic radiation represented by vector potential **A** with an electron in potential $V$ is given by [10,32]

$$H_{int} = \frac{e}{m_e c}\mathbf{A}\cdot\mathbf{p} - \frac{i\hbar e}{2m_e c}\nabla\cdot\mathbf{A} + \frac{e\hbar}{4m_e^2 c^3}\boldsymbol{\sigma}\times\nabla V\cdot\mathbf{A} \quad , \qquad (3)$$

The three terms correspond to momentum-conserving dipole transition, surface photoemission, and spin-orbit coupling. The transition matrix element, after accounting for the dielectric discontinuity at the surface, becomes

$$\begin{aligned}\langle\psi_f|H_{int}|\psi_i\rangle \propto &\frac{e}{m_e c}\left\langle\psi_f\left|A_x p_x + A_y p_y + \frac{1}{\varepsilon}A_z p_z\right|\psi_i\right\rangle \\ &-\frac{i\hbar e}{m_e c}\left(1-\frac{1}{\varepsilon}\right)\pi\psi_f^*(0)\psi_i(0)A_z + \beta\langle\psi_f|(A_x\sigma_y - A_y\sigma_x)|\psi_i\rangle\end{aligned} \quad , \qquad (4)$$

where $\varepsilon$ is the dielectric constant of Bi$_2$Te$_3$, and $\beta = 4.32\times10^{-3}\ e$ is a Rashba parameter determined by the Dirac cone dispersion or energy splitting as a function of momentum, which is available from first-principles calculations or from experimental dispersions [23].

First-principles calculations yield the initial state $\psi_i$. The final state $\psi_f$ is a time-reversed low-energy-electron-diffraction (TRLEED) state [24,25], keeping only the zeroth order of the crystal potential (the inner potential 8.5 eV) and accounting for multiple reflections and damping of the electron waves within the film [23]. The damping (or decoherence) length equals twice the photoelectron mean free path and is taken to be 12 Å. The reflection coefficient at the interface, taken to be a linear function of electron energy, is extracted from fitting to the data. The result corresponds to an intensity reflectivity around 70%, which mainly affects the 1 QL case only because the damping suppresses multiple reflections at larger film thicknesses. The only other parameter needed for the computation is the complex dielectric constant $\varepsilon$ of $Bi_2Te_3$, which should be close to unity in the energy range of interest [33]. We treat $\varepsilon$ as a fitting parameter that depends on the photon energy but not on the film thickness. It does come out to be close to unity [23]. The calculation also includes contributions from the buried interface, which are relatively minor because of the damping of the final state. The results of the calculation (Fig. 2(c)) are in overall agreement with the data. The minor differences can be attributed to the various approximations. A further comparison of the CD involving the $k_y$ dependence of the upper branch in selected cases is presented in Fig. 3. The CD is not an odd function, as expected. Note the large differences between the results at 55 and 60 eV (sign reversal), and the unusual behavior for 1 QL at 55 eV, are all reproduced by the calculation. The detailed agreement lends strong support to our analysis. Note that even the "bulk" bands can exhibit a nontrivial CD due to the strong spin-orbit coupling, but the effects are generally weaker compared to the topological surface states.

Prior discussions of CD [15,16,34] generally invoked only the dipole term in Eq. (3). Is this justified? We show in Figs. 4(a)-(c) the calculated CD with only two of the three terms in Eq. (3) included and in Figs. 4(d)-(f) with only one of the three terms included. Note that the matrix elements are complex functions. Interference of the three contributions makes the CD values large in Figs. 4(a) and 4(b), but quite small in the other cases. A single-term description yields very small CD and is simply not adequate. With two terms, only the case including both the dipole term and the spin-orbit term comes roughly close to the experiment. Thus, the contribution of the surface photoemission term is relatively weak but not negligible. In general, the surface photoemission term is weak relative to the dipole term if direct transitions are allowed, and vice-versa [32]. The $c$-axis lattice constant of $Bi_2Te_3$ is very large [22], leading to a very small dimension of the Brillouin zone along this direction. The final state is dominated by a free-electron state folded into the narrow first zone. Direct band-to-band transition is essentially continuously allowed, independent of the photon energy. Thus, the surface photoemission term is expected to be weaker.

Additional data based on geometry B reveal an odd CD function as expected from symmetry requirements, and the results are again in good agreement with the calculation [23]. Some insights can be garnered from the analysis. In general, CD measurements at just one randomly chosen photon energy or just one film thickness are not necessarily a straightforward indication of the spin polarization of the topological state, and the sign can be reversed. At high photon energies where $\varepsilon$ approaches unity or in cases direct transitions are allowed, the surface photoemission term can be relatively weak. The spin-orbit term is generally strong for systems with heavy elements. At very low photon energies (such as 6 eV), where the final state damping effect is suppressed, the contributions from the top and bottom faces of the film, with opposite spin textures, can interfere with each other. In systems with a typical lattice constant along the $c$-direction, direct transitions are strongly modulated as a function of photon energy and can be strategically minimized. These guidelines are helpful for simplifying the analysis of CD in terms of the spin polarization by choosing appropriate experimental conditions.

In summary, we have performed a comprehensive study of the CD of the topological surface states in ultrathin films of $Bi_2Te_3$ grown on Si(111). The experimentally observed complex dependencies of CD on the photon energy, film thickness, crystal momentum, and experimental geometry (or initial states with different symmetries) are all well described by the calculation. Our results resolve the outstanding issues related to the apparent complexity of experimental CD results. While the spin polarization of the topological surface states plays an important role in the analysis, extraction of this quantity requires a thorough understanding of the photoemission process. While it is expected that circularly polarized light, which carries an angular momentum, can be used to manipulate the spin of photoelectrons [34–36], our study shows that the photoelectron spin can be further controlled by changing the film thickness or photon energy. This control offers opportunities for photocathode applications. The same analysis and methodology should be broadly applicable to surfaces and thin films in general.

This work is supported by the U.S. Department of Energy, Office of Science, Office of Basic Energy Sciences, under Grant No. DE-FG02-07ER46383 (TCC). We thank M. Bissen and M. Severson for assistance with beamline operation at the Synchrotron Radiation Center, which was supported by the University of Wisconsin-Madison. T. Miller and the beamline operations were partially supported by NSF Grant No. DMR 13-05583.

# References

[1]   L. Fu, C. Kane, and E. Mele, Phys. Rev. Lett. **98**, 106803 (2007).

[2]   J. Moore and L. Balents, Phys. Rev. B **75**, 121306 (2007).

[3]   R. Roy, Phys. Rev. B **79**, 195322 (2009).

[4]   M. Z. Hasan and C. L. Kane, Rev. Mod. Phys. **82**, 3045 (2010).

[5]   H. Zhang, C.-X. Liu, X.-L. Qi, X. Dai, Z. Fang, and S.-C. Zhang, Nat. Phys. **5**, 438 (2009).

[6]   J. E. Moore, Nature **464**, 194 (2010).

[7]   Y. S. Kim, M. Brahlek, N. Bansal, E. Edrey, G. A. Kapilevich, K. Iida, M. Tanimura, Y. Horibe, S. W. Cheong, and S. Oh, Phys. Rev. B **84**, 073109 (2011).

[8]   H. Z. Lu, W. Y. Shan, W. Yao, Q. Niu, and S. Q. Shen, Phys. Rev. B **81**, 115407 (2010).

[9]   M. Neupane, A. Richardella, J. Sánchez-Barriga, S. Xu, N. Alidoust, I. Belopolski, C. Liu, G. Bian, D. Zhang, D. Marchenko, A. Varykhalov, O. Rader, M. Leandersson, T. Balasubramanian, T.-R. Chang, H.-T. Jeng, S. Basak, H. Lin, A. Bansil, N. Samarth, and M. Z. Hasan, Nat. Commun. **5**, 4841 (2014).

[10]  Y. Wang and N. Gedik, Phys. Status Solidi RRL **7**, 64 (2013).

[11]  J. Stöhr and H. C. Siegmann, *Magnetism: From Fundamentals to Nanoscale Dynamics* (Springer Berlin, 2006).

[12]  Y. H. Wang, D. Hsieh, D. Pilon, L. Fu, D. R. Gardner, Y. S. Lee, and N. Gedik, Phys. Rev. Lett. **107**, 207602 (2011).

[13]  Y. Ishida, H. Kanto, A. Kikkawa, Y. Taguchi, Y. Ito, Y. Ota, K. Okazaki, W. Malaeb, M. Mulazzi, M. Okawa, S. Watanabe, C.-T. Chen, M. Kim, C. Bell, Y. Kozuka, H. Y. Hwang, Y. Tokura, and S. Shin, Phys. Rev. Lett. **107**, 077601 (2011).

[14]  S. R. Park, J. Han, C. Kim, Y. Y. Koh, C. Kim, H. Lee, H. J. Choi, J. H. Han, K. D. Lee, N. J. Hur, M. Arita, K. Shimada, H. Namatame, and M. Taniguchi, Phys. Rev. Lett. **108**, 046805 (2012).

[15]  H. Mirhosseini and J. Henk, Phys. Rev. Lett. **109**, 036803 (2012).

[16] M. R. Scholz, J. Sánchez-Barriga, J. Braun, D. Marchenko, A. Varykhalov, M. Lindroos, Y. J. Wang, H. Lin, A. Bansil, J. Minár, H. Ebert, A. Volykhov, L. V. Yashina, and O. Rader, Phys. Rev. Lett. **110**, 216801 (2013).

[17] F. Vidal, M. Eddrief, B. Rache Salles, I. Vobornik, E. Velez-Fort, G. Panaccione, and M. Marangolo, Phys. Rev. B **88**, 241410(R) (2013).

[18] Y.-Y. Li, G. Wang, X.-G. Zhu, M.-H. Liu, C. Ye, X. Chen, Y.-Y. Wang, K. He, L.-L. Wang, X.-C. Ma, H.-J. Zhang, X. Dai, Z. Fang, X.-C. Xie, Y. Liu, X.-L. Qi, J.-F. Jia, S.-C. Zhang, and Q.-K. Xue, Adv. Mater. **22**, 4002 (2010).

[19] Y. Liu, G. Bian, T. Miller, M. Bissen, and T.-C. Chiang, Phys. Rev. B **85**, 195442 (2012).

[20] X. Gonze, B. Amadon, P. M. Anglade, J. M. Beuken, F. Bottin, P. Boulanger, F. Bruneval, D. Caliste, R. Caracas, M. Côté, T. Deutsch, L. Genovese, P. Ghosez, M. Giantomassi, S. Goedecker, D. R. Hamann, P. Hermet, F. Jollet, G. Jomard, S. Leroux, M. Mancini, S. Mazevet, M. J. T. Oliveira, G. Onida, Y. Pouillon, T. Rangel, G. M. Rignanese, D. Sangalli, R. Shaltaf, M. Torrent, M. J. Verstraete, G. Zerah, and J. W. Zwanziger, Comput. Phys. Commun. **180**, 2582 (2009).

[21] C. Hartwigsen, S. Goedecker, and J. Hutter, Phys. Rev. B **58**, 3641 (1998).

[22] W. Zhang, R. Yu, H.-J. Zhang, X. Dai, and Z. Fang, New J. Phys. **12**, 065013 (2010).

[23] See Supplemental Material [url] for additional data and details of the model, which includes Refs. [24-27].

[24] B. Feuerbacher, B. Fitton, and R. F. Willis, *Photoemission and the Electronic Properties of Surfaces* (Wiley, New York, 1978).

[25] G. Mahan, Phys. Rev. B **2**, 4334 (1970).

[26] C. X. Liu, X. L. Qi, H. Zhang, X. Dai, Z. Fang, and S. C. Zhang, Phys. Rev. B **82**, 045122 (2010).

[27] A. Herdt, L. Plucinski, G. Bihlmayer, G. Mussler, S. Döring, J. Krumrain, D. Grützmacher, S. Blügel, and C. Schneider, Phys. Rev. B **87**, 035127 (2013).

[28] Y. Zhang, K. He, C.-Z. Chang, C.-L. Song, L.-L. Wang, X. Chen, J.-F. Jia, Z. Fang, X. Dai, W.-Y. Shan, S.-Q. Shen, Q. Niu, X.-L. Qi, S.-C. Zhang, X.-C. Ma, and Q.-K. Xue, Nat. Phys. **6**, 584 (2010).

[29]  O. V. Yazyev, J. E. Moore, and S. G. Louie, Phys. Rev. Lett. **105**, 266806 (2010).

[30]  X. Wang, G. Bian, T. Miller, and T. C. Chiang, Phys. Rev. Lett. **108**, 096404 (2012).

[31]  Z. H. Zhu, C. N. Veenstra, S. Zhdanovich, M. P. Schneider, T. Okuda, K. Miyamoto, S. Y. Zhu, H. Namatame, M. Taniguchi, M. W. Haverkort, I. S. Elfimov, and A. Damascelli, Phys. Rev. Lett. **112**, 076802 (2014).

[32]  T. Miller, W. McMahon, and T.-C. Chiang, Phys. Rev. Lett. **77**, 1167 (1996).

[33]  D. Attwood, *Soft X-Rays and Extreme Ultraviolet Radiation: Principles and Applications* (Cambridge University Press, Cambridge, 1999).

[34]  J. Sánchez-Barriga, A. Varykhalov, J. Braun, S.-Y. Xu, N. Alidoust, O. Kornilov, J. Minár, K. Hummer, G. Springholz, G. Bauer, R. Schumann, L. V. Yashina, H. Ebert, M. Z. Hasan, and O. Rader, Phys. Rev. X **4**, 011046 (2014).

[35]  C. H. Park and S. G. Louie, Phys. Rev. Lett. **109**, 097601 (2012).

[36]  C. Jozwiak, C.-H. Park, K. Gotlieb, C. Hwang, D.-H. Lee, S. G. Louie, J. D. Denlinger, C. R. Rotundu, R. J. Birgeneau, Z. Hussain, and A. Lanzara, Nat. Phys. **9**, 293 (2013).

FIG. 1 (color online). (a)-(b) ARPES geometries A and B. Green hexagons represent the surface Brillouin zone of $Bi_2Te_3$. (c)-(d) ARPES maps of a 2-QL $Bi_2Te_3$ film taken with LCP and RCP light at 55 eV, respectively, using geometry A. The brightness indicates the photoemission intensity. (e)-(f) ARPES maps of 2-QL and 12-QL $Bi_2Te_3$ films, respectively, taken with linearly polarized light at 29 eV.

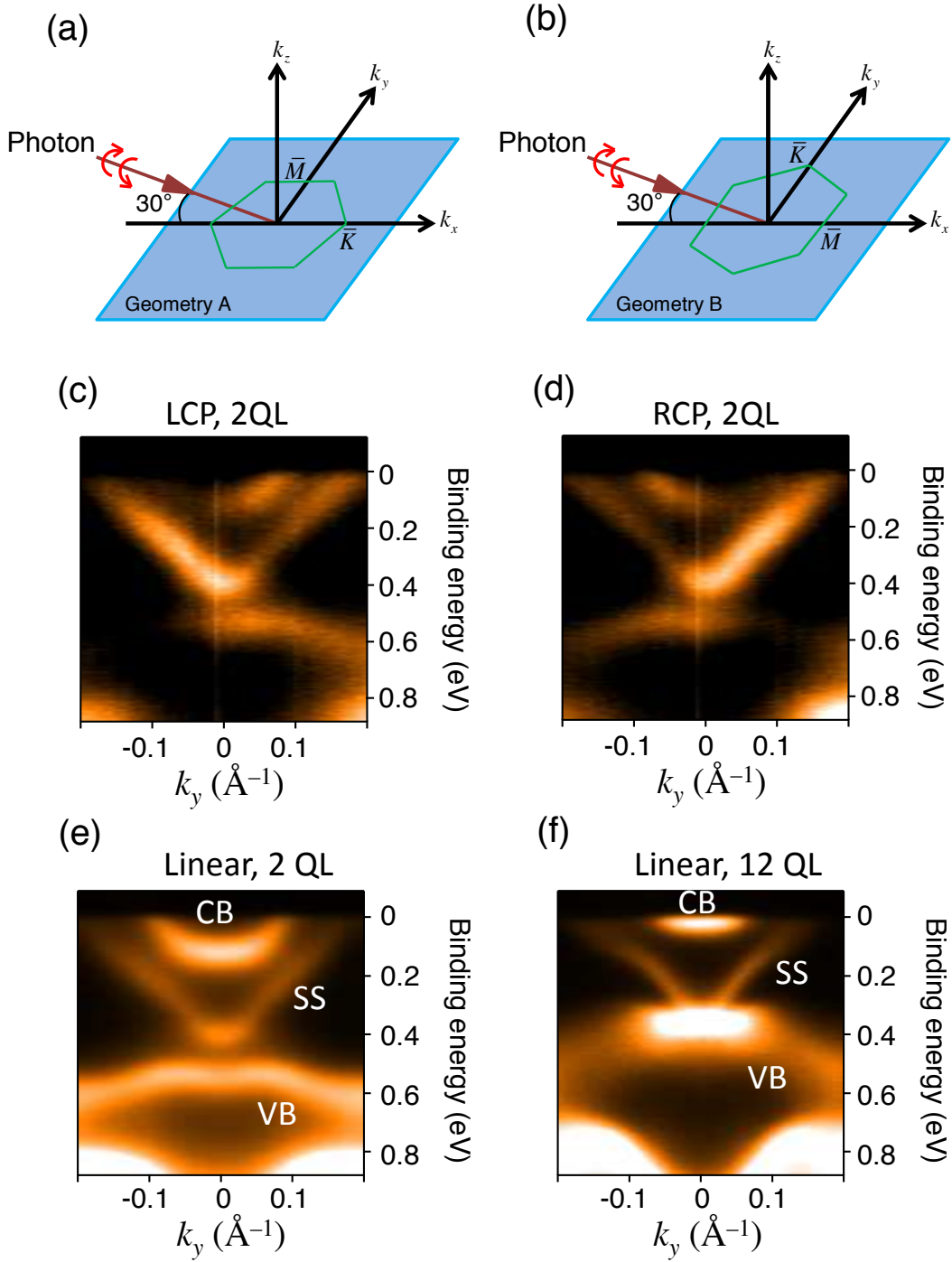

FIG. 2 (color online). Circular dichroism of $Bi_2Te_3$ films, with a comparison between experiment and model calculation. (a) Maps of photoemission CD for films of $Bi_2Te_3$ of various thicknesses taken at different photon energies using geometry A. The green dashed curves indicate the dispersion of the upper topological surface state. (b) Photon energy and film thickness dependences of CD for the upper topological surface state at $k_y = -0.08$ Å$^{-1}$. (c) Computed CD for comparison with results shown in (b). (d) Calculated spin polarization of the upper topological state at $k_y = -0.08$ Å$^{-1}$.

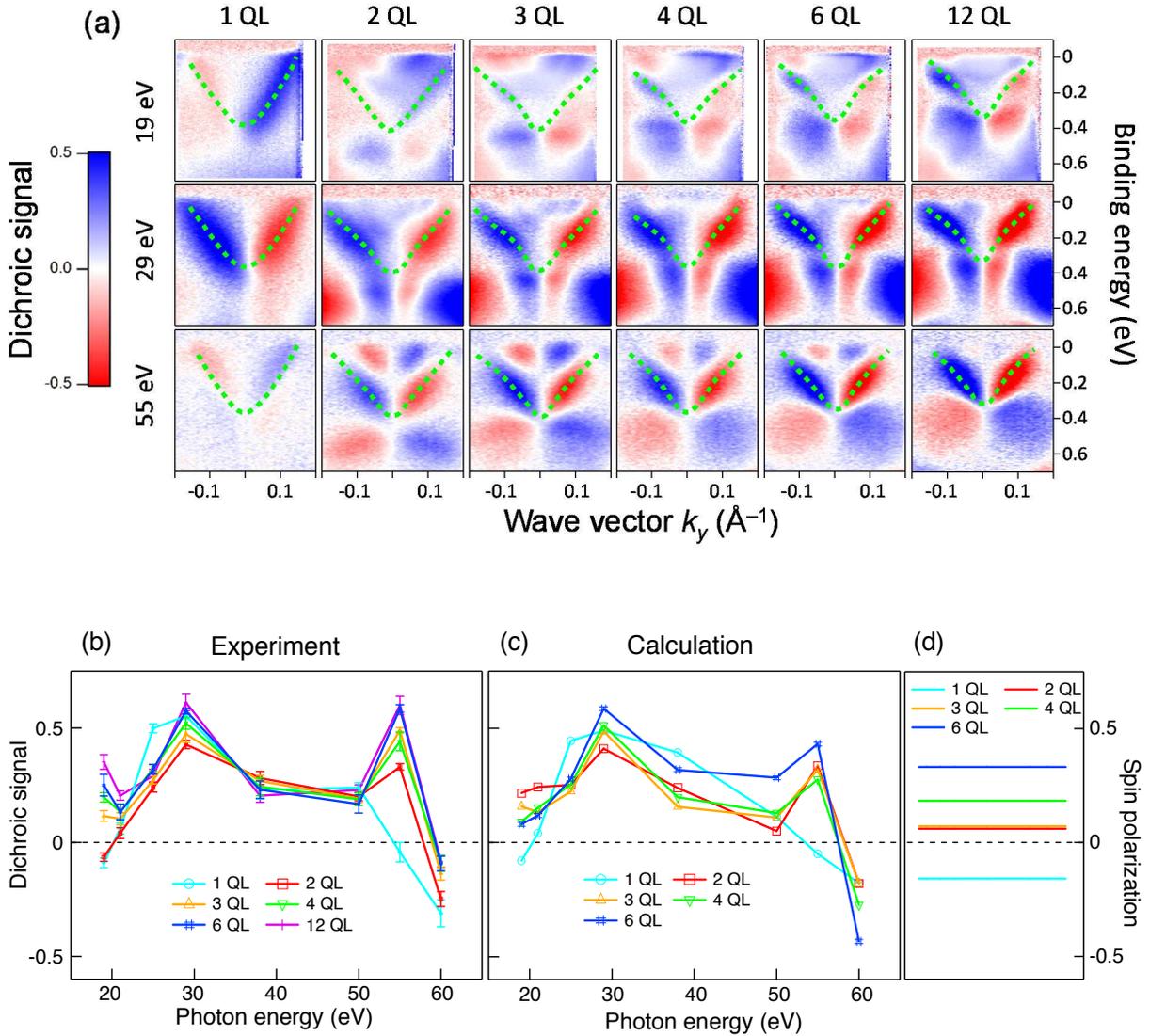

FIG. 3 (color online). Experimental and theoretical CD of $Bi_2Te_3$ films of various thicknesses taken at different photon energies using geometry A. The results are for the upper branch of the topological states as a function of $k_y$. The decreasing CD signal at larger $k_y$ compared to the calculation is caused by overlapping signal from the neighboring conduction band states because of a finite band width in the experiment. Likewise, the sharp changes in theoretical CD around the zone center, especially at 29 and 55 eV, become rounded in the experiment due overlapping signal from the neighboring valence band states.

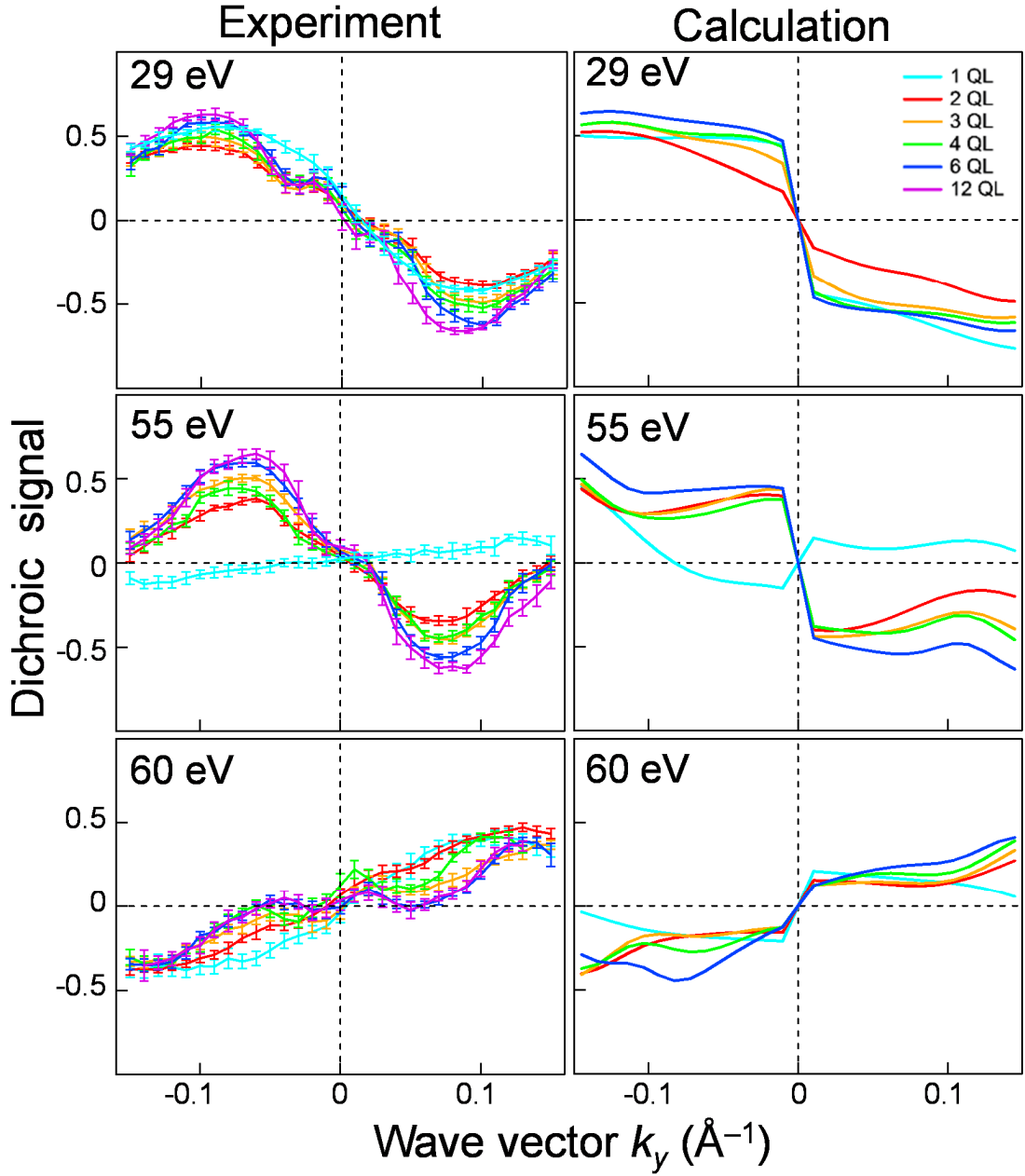

FIG. 4 (color online). Computed CD for geometry A with the following reduced set of contributions of photoemission: (a) No dipole transition (DT); (b) no surface photoemission (SP); (c) no spin-orbit coupling (SOC); (d) dipole transition only; (e) surface photo-emission only; and (f) spin-orbit coupling only. The CD is computed for the upper topological state at $k_y = -0.08$ Å$^{-1}$, the same wave vector as that in Fig. 2(c).

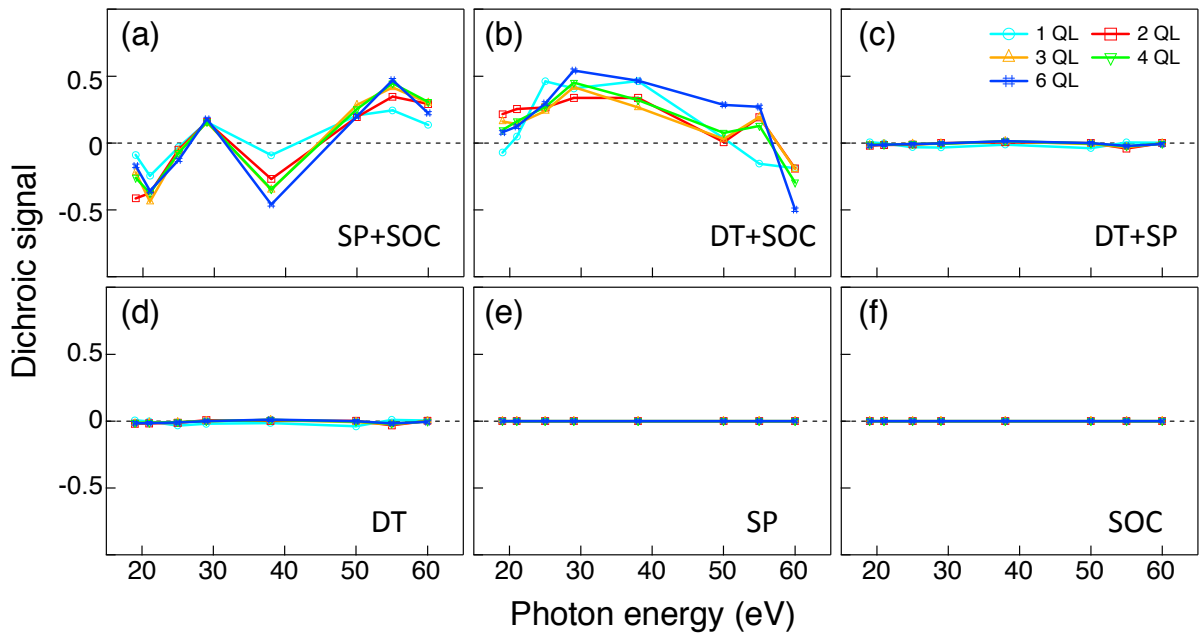